\documentclass[aps,prl,showpacs,floatfix,twocolumn,byrevtex,superscriptaddress]{revtex4-1}
\usepackage{amsmath}
\usepackage{amssymb}
\usepackage{amstext}
\usepackage{amsopn}
\usepackage{amsfonts}
\usepackage{amsxtra}
\usepackage[english]{babel}
\usepackage{graphicx}
\usepackage{float}
\usepackage{bm}
\usepackage{multirow}
\usepackage{dcolumn}
\usepackage{color}
\usepackage{hyperref}
\usepackage[normalem]{ulem}

\newcommand{\icm}{\ensuremath{\textrm{cm}^{-1}}}

\newcommand{\LNOinf}{LaNiO$_{3}$}
\newcommand{\LNObi}{La$_{3}$Ni$_{2}$O$_{7}$}
\newcommand{\LNOtri}{La$_{4}$Ni$_{3}$O$_{10}$}

\newcommand{\Kratio}{$K_{\text{exp}}/K_{\text{band}}$}

\begin{document}

\title{Evolution of Electronic Correlations in the Ruddlesden-Popper Nickelates}
\author{Zhe~Liu}
\thanks{These authors contributed equally to this work.}
\author{Jie~Li}
\thanks{These authors contributed equally to this work.}
\affiliation{National Laboratory of Solid State Microstructures and Department of Physics, Collaborative Innovation Center of Advanced Microstructures, Nanjing University, Nanjing 210093, China}
\author{Mengwu~Huo}
\thanks{These authors contributed equally to this work.}
\affiliation{Center for Neutron Science and Technology, Guangdong Provincial Key Laboratory of Magnetoelectric Physics and Devices, School of Physics, Sun Yat-Sen University, Guangzhou, Guangdong 510275, China}
\author{Bingke~Ji}
\author{Jiahao~Hao}
\author{Yaomin~Dai}
\email{ymdai@nju.edu.cn}
\author{Mengjun~Ou}
\author{Qing~Li}
\affiliation{National Laboratory of Solid State Microstructures and Department of Physics, Collaborative Innovation Center of Advanced Microstructures, Nanjing University, Nanjing 210093, China}
\author{Hualei~Sun}
\affiliation{School of Sciences, Sun Yat-Sen University, Shenzhen, Guangdong 518107, China}
\author{Bing~Xu}
\affiliation{Beijing National Laboratory for Condensed Matter Physics, Institute of Physics, Chinese Academy of Sciences, P.O. Box 603, Beijing 100190, China}
\author{Yi~Lu}
\email{yilu@nju.edu.cn}
\affiliation{National Laboratory of Solid State Microstructures and Department of Physics, Collaborative Innovation Center of Advanced Microstructures, Nanjing University, Nanjing 210093, China}
\author{Meng~Wang}
\email{wangmeng5@mail.sysu.edu.cn}
\affiliation{Center for Neutron Science and Technology, Guangdong Provincial Key Laboratory of Magnetoelectric Physics and Devices, School of Physics, Sun Yat-Sen University, Guangzhou, Guangdong 510275, China}
\author{Hai-Hu~Wen}
\email{hhwen@nju.edu.cn}
\affiliation{National Laboratory of Solid State Microstructures and Department of Physics, Collaborative Innovation Center of Advanced Microstructures, Nanjing University, Nanjing 210093, China}

\date{\today}
%
%

\begin{abstract}
We report on optical studies of the Ruddlesden-Popper nickelates La$_{n+1}$Ni$_{n}$O$_{3n+1}$ with $n = 2$ (La$_{3}$Ni$_{2}$O$_{7}$), $n = 3$ (La$_{4}$Ni$_{3}$O$_{10}$) and $n = \infty$ (LaNiO$_{3}$). As the number of the NiO$_{6}$ octahedra layers $n$ grows, the ratio of the kinetic energy determined from the experimental optical conductivity and that from band theory $K_{\text{exp}}/K_{\text{band}}$ increases, suggesting a reduction of electronic correlations. While the strong electronic correlations in the bilayer La$_{3}$Ni$_{2}$O$_{7}$ place it on the verge of the Mott insulating phase, the trilayer La$_{4}$Ni$_{3}$O$_{10}$ and infinite-layer LaNiO$_{3}$ exhibit moderate electronic correlations, falling into the regime of correlated metals. The evolution of the electronic correlations in La$_{n+1}$Ni$_{n}$O$_{3n+1}$ is likely to be dominated by the Ni-$d_{z^2}$ orbital. Our results provide important information for understanding the superconductivity in Ruddlesden-Popper nickelates.
\end{abstract}



\maketitle

%
%
In the past two decades, the search for high critical temperature (high-$T_{c}$) superconductivity (SC) in nickelates has been following the strategy of mimicking the high-$T_{c}$ cuprates~\cite{Anisimov1999PRB}: embedding Ni$^{+}$ with a $3d^{9}$ electronic configuration into a square planar coordination of O ions to mimic the parent compound of cuprate such as CaCuO$_{2}$ which is a Mott insulator, and then doping low-spin holes into the NiO$_{2}$ planes. This strategy has led to the realization of SC with a $T_{c}$ up to 18.8~K in the hole-doped infinite-layer nickelates $Re_{1-x}A_{x}$NiO$_{2}$ ($Re$ = La, Pr, Nd; $A$ = Sr, Ca)~\cite{Li2019Nature,Li2020PRL,Zeng2020PRL,Osada2020NL,Osada2021AM,Sun2023AM,Zeng2022SA} and the stoichiometric quintuple-layer Nd$_{6}$Ni$_{5}$O$_{12}$ thin films~\cite{Pan2022NM}. The onset $T_{c}$ of the Pr$_{0.82}$Sr$_{0.18}$NiO$_{2}$ thin films can be enhanced monotonically from 17~K at ambient pressure to 31~K at 12.1~GPa without showing any trend towards saturation~\cite{Wang2022NC}. However, SC thus far has not been observed in bulk infinite-layer nickelates~\cite{Li2020CM,Wang2020PRM,Huo2022CPB}.

The recent discovery of SC with a $T_{c}$ up to 80~K in the bilayer \LNObi\ single crystal under pressure has opened a new avenue towards high-$T_{c}$ SC in nickelates~\cite{Sun2023Nature,Wang2024PRX,Zhang2024NP}. Shortly afterwards, the trilayer \LNOtri\ has also been found to exhibit SC with a $T_{c}$ up to 30~K under high pressure~\cite{Zhu2024Nature,Li2024CPL,Zhang2023arXiv,Li2024SCPMA,Huang2024arXiv}. These two compounds belong to the Ruddlesden-Popper (RP) phase nickelates La$_{n+1}$Ni$_{n}$O$_{3n+1}$, which are characterized by $n$ consecutive layers of corner-sharing NiO$_{6}$ octahedra between rock salt LaO layers. Both \LNObi\ and \LNOtri\ exhibit charge/spin density-wave transitions at ambient pressure~\cite{Chen2024PRL,Zhang2020NC,Liu2023SCPMA,Chen2024arXiv,Dan2024arXiv}. The emergence of SC is accompanied by the suppression of the density-wave ordering, and strange metal behavior is observed above the SC dome in the temperature-pressure ($T$-$p$) phase diagram~\cite{Sun2023Nature,Wang2024PRX,Zhang2024NP,Zhu2024Nature,Li2024CPL,Zhang2023arXiv}. Spectroscopic studies have revealed strong electronic correlation effects in these RP nickelates~\cite{Li2017NC,Yang2024NC,Liu2024NC,Fan2024PRB,Abadi2024arXiv,Li2024arXiv,Du2024arXiv}. Since theoretical work has underlined the important role of electronic correlations in driving unconventional SC in the RP phase nickelates~\cite{Lechermann2023PRB,Liao2023PRB,Sakakibara2024PRL,Zhang2024PRL,Lu2024PRL,Yang2023PRB,Jiang2024CPL,Qu2024PRL,Oh2023PRB}, it would be highly instructive to establish the evolution of electronic correlations across the RP nickelate family and unveil the underlying factors that control the electronic correlations.

In this paper, we investigate the evolution of electronic correlations in the RP nickelates La$_{n+1}$Ni$_{n}$O$_{3n+1}$ using optical spectroscopy. As $n$ grows, the kinetic energy ratio $K_{\text{exp}}/K_{\text{band}}$ increases, indicating a reduction of electronic correlations with increasing NiO$_{6}$ layers. While the bilayer La$_{3}$Ni$_{2}$O$_{7}$ exhibits strong electronic correlations which place it in the proximity to the Mott insulating phase, the trilayer La$_{4}$Ni$_{3}$O$_{10}$ and infinite-layer LaNiO$_{3}$ fall into the correlated metal regime, characterized by moderate electronic correlations. Further analysis implies that the evolution of electronic correlations in RP nickelates is governed by the Ni-$d_{z^{2}}$ orbital. The simultaneous suppression of electronic correlations and the maximum $T_{c}$ with $n$ hints that electronic correlations may play an important role in promoting high-$T_{c}$ SC in RP nickelates.

%
%

%
%

\begin{figure*}[tb]
\includegraphics[width=\textwidth]{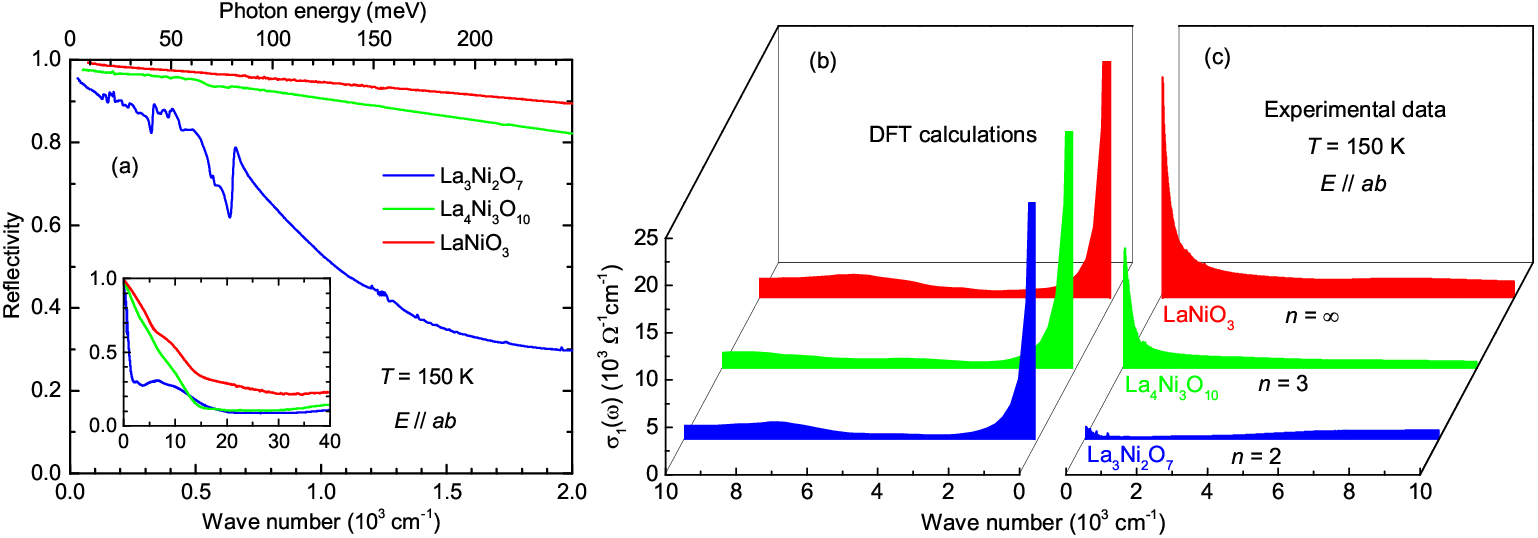}
\caption{(a) $R(\omega)$ of \LNObi\ (blue curve), \LNOtri\ (green curve), and \LNOinf\ (red curve) up to 2000~\icm\ at 150~K. The inset shows $R(\omega)$ of the three compounds in a broader frequency range up to 40\,000~\icm. (b) and (c) show the calculated and experimental $\sigma_{1}(\omega)$ of \LNObi\ (blue shaded area), \LNOtri\ (green shaded area) and \LNOinf\ (red shaded area), respectively.}
\label{ERefS1}
\end{figure*}
Sample characterizations and experimental details are given in the Supplementary Materials~\cite{SuppMat} and Refs.~\cite{Homes1993,Dressel2002,Tanner2019}. Figure~\ref{ERefS1}(a) displays the reflectivity $R(\omega)$ of \LNObi\ (blue solid curve), \LNOtri\ (green solid curve), and \LNOinf\ (red solid curve) at 150~K for $E \parallel ab$. The $R(\omega)$ spectra for all three compounds approach unity in the low-frequency limit, suggesting that \LNObi, \LNOtri, and \LNOinf\ are all metallic. However, $R(\omega)$ of \LNObi\ is relatively low and characterized by IR-active phonon modes (sharp features) in the far-infrared range, implying very poor metallicity. With increasing NiO$_{6}$ layers, such as \LNOtri\ ($n = 3$) and \LNOinf\ ($n = \infty$), the far-infrared $R(\omega)$ increases and the IR-active phonon modes are screened, indicating enhanced metallicity. The inset of Fig.~\ref{ERefS1}(a) shows $R(\omega)$ of the three materials up to 40\,000~\icm. A plasma edge is observed for all three compounds. As $n$ increases, the plasma edge shifts to higher frequency, which also attests to an enhancement of metallicity.

Figure~\ref{ERefS1}(c) shows the experimental optical conductivity $\sigma_{1}(\omega)$ of \LNObi\ (blue shaded area), \LNOtri\ (green shaded area), and \LNOinf\ (red shaded area) at 150~K. All three compounds exhibit a Drude response, i.e. a peak centered at zero frequency, in the low-frequency $\sigma_{1}(\omega)$ spectra, which is the distinguishing characteristic of a metal. While the Drude weight [the area under the Drude profile in $\sigma_{1}(\omega)$] is vanishingly small for \LNObi, it grows significantly as $n$ increases, suggesting an enhancement of metallicity. The Drude weight is proportional to the electron's kinetic energy defined as~\cite{Millis2005PRB,Qazilbash2009NP}
%
%
\begin{equation}
K = \frac{2\hbar^{2} c_{0}}{\pi e^2}\int_{0}^{\omega_{c}}\sigma_{1}(\omega)\text{d}\omega,
\label{Kinetic}
\end{equation}
where $c_{0}$ is the $c$-axis lattice parameter, and $\omega_{c}$ is a cutoff frequency which should be high enough to cover the entire Drude component in $\sigma_{1}(\omega)$ but not so high as to include considerable interband contributions. In correlated materials, electronic correlations impede the motion of electrons, leading to a reduction of the electron's kinetic energy, i.e. Drude weight, compared to the prediction of band theory which does not take electronic correlations into account. Hence, a comparison of the Drude weight in the experimental $\sigma_{1}(\omega)$ and that in the calculated $\sigma_{1}(\omega)$ from density-functional theory (DFT) conveniently reflects the strength of electronic correlations~\cite{Millis2005PRB,Qazilbash2009NP,Si2009NP}.

Figure~\ref{ERefS1}(b) displays the calculated $\sigma_{1}(\omega)$ for \LNObi\ (blue shaded area), \LNOtri\ (green shaded area), and \LNOinf\ (red shaded area) from DFT. The details about the DFT calculations can be found in the Supplementary Materials~\cite{SuppMat} and Refs.~\cite{Blaha2001,Perdew1996PRL}. For \LNObi, the Drude weight in the experimental $\sigma_{1}(\omega)$ [blue shaded area in Fig.~\ref{ERefS1}(c)] is remarkably smaller than that in the calculated $\sigma_{1}(\omega)$ [blue shaded area in Fig.~\ref{ERefS1}(b)], indicating very strong electronic correlations in this compound~\cite{Liu2024NC}. As $n$ grows, the difference in the Drude weight between the experimental and calculated $\sigma_{1}(\omega)$ diminishes, unambiguously pointing to a reduction of electronic correlations with increasing NiO$_{6}$ layers.

\begin{figure*}[tb]
\includegraphics[width=\textwidth]{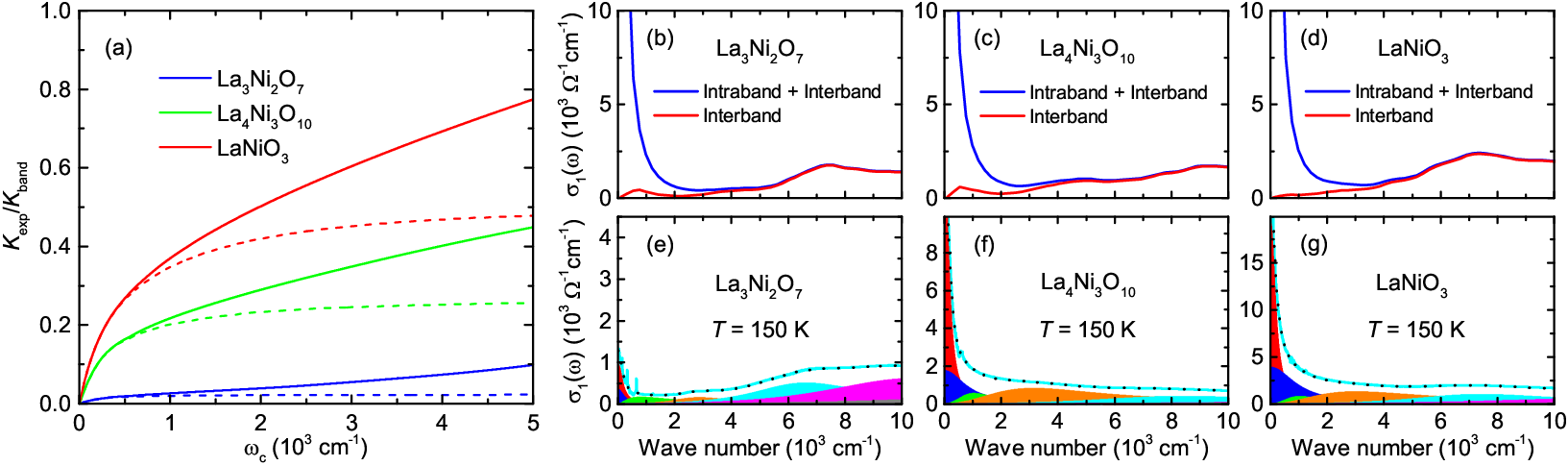}
\caption{(a) The Kinetic energy ratio \Kratio\ as a function of the cutoff frequency $\omega_{c}$ for \LNObi\ (blue curves), \LNOtri\ (green curves) and \LNOinf\ (red curves). The dashed lines denote \Kratio\ with the interband contributions removed. (b)-(d) Calculated $\sigma_{1}(\omega)$ for \LNObi, \LNOtri\ and \LNOinf. The blue solid curve in each panel denotes $\sigma_{1}(\omega)$ including intraband contributions. (e)-(g) The cyan solid curve in each panel is the measured $\sigma_{1}(\omega)$ at 150~K. The black dashed line through the data represents the Drude-Lorentz fit which is decomposed into two Drude components (red and blue shaded areas) and a series of Lorentz components (shaded areas in other colors).}
\label{EFit}
\end{figure*}
The ratio of the experimental kinetic energy and the kinetic energy from band theory \Kratio\ represents a more quantitative measure of electronic correlations. This approach has been adopted to explore the electronic correlations in many different systems including cuprates~\cite{Millis2005PRB}, iron pnictides~\cite{Qazilbash2009NP,Schafgans2012PRL,Degiorgi2011NJP}, topological semimetals~\cite{Shao2020NP} and kagome metals~\cite{Xu2020NC,Zhou2023PRBAVS}. $K_{\text{exp}}$ can be determined from the integral of the measured $\sigma_{1}(\omega)$ using Eq.~(\ref{Kinetic}), and $K_{\text{band}}$ can be directly obtained from the DFT calculations~\cite{SuppMat}. Figure~\ref{EFit}(a) depicts \Kratio\ as a function of the cutoff frequency $\omega_{c}$ for \LNObi\ (blue solid curve), \LNOtri\ (green solid curve), and \LNOinf\ (red solid curve). For all three compounds, \Kratio\ grows monotonically as $\omega_{c}$ goes up. This is caused by the inclusion of a sizeable amount of contributions from interband transitions in the high-frequency range. Nevertheless, it is noteworthy that regardless of the choice of $\omega_{c}$, \Kratio\ increases as the NiO$_{6}$ layers increase, corroborating the reduction of electronic correlations with increasing NiO$_{6}$ layers.

The calculated $\sigma_{1}(\omega)$ [Figs.~\ref{EFit}(b)-\ref{EFit}(d)] manifests that interband transitions are present in the low-energy range for all three compounds, resulting in a considerable overlap between intraband and interband excitations in the $\sigma_{1}(\omega)$ spectra. Therefore, an accurate determination of \Kratio\ requires the subtraction of interband contributions from $\sigma_{1}(\omega)$. To separate the intraband and interband contributions, we fit the measured $\sigma_{1}(\omega)$ to the Drude-Lorentz model~\cite{Dressel2002,Tanner2019},
%
%
\begin{equation}
\sigma_{1}(\omega) = \frac{2\pi}{Z_{0}} \left[\sum_{k} \frac{\omega^{2}_{p,k}}{\tau_{k}(\omega^{2}+\tau_{k}^{-2})}
   + \sum_{i} \frac{\gamma_{i} \omega^{2} \omega_{p,i}^{2}}{(\omega_{0,i}^{2} - \omega^{2})^{2} + \gamma_{i}^{2} \omega^{2}}\right],
\label{DLModel}
\end{equation}
where $Z_{0} \simeq 377$~$\Omega$ refers to the impedance of free space. The Drude model (the first term) describes the optical response of intraband transitions, which is characterized by a plasma frequency $\omega_{p}$ and a quasiparticle scattering rate $1/\tau$; the second term is a sum of Lorentzian oscillators which are used to model interband transitions. Here, $\omega_{0,i}$, $\gamma_{i}$, and $\omega_{p,i}$ are the position, line width, and strength of the $i$th oscillator.

Figures~\ref{EFit}(e)-\ref{EFit}(g) show that the measured $\sigma_{1}(\omega)$ for all three compounds (cyan solid curves) can be fit quite well by the Drude-Lorentz model, as depicted by the black dashed lines through the data, which consists of two Drude components (red and blue shaded areas) and a series of Lorentz components (shaded areas in other colors). The utilization of two Drude components is consistent with the existence of multiple bands arising from different Ni-$3d$ orbitals at the Fermi level in the RP nickelates~\cite{Sun2023Nature,Liu2024NC,Nakata2017PRB,Zhang2023PRB,Wang2024PRB,Leonov2024PRB,Yang2024PRB,Chen2024PRB,Malashevich2015PRB}. The discrepancy in the position and weight of interband transitions between the experimental and calculated $\sigma_{1}(\omega)$ may result from band renormalization due to electronic correlations~\cite{Qazilbash2009NP,Si2009NP,Xu2020NC,Geisler2024arXiv}. We subtract the interband contributions from the experimental $\sigma_{1}(\omega)$ and obtain \Kratio\ as a function of $\omega_{c}$ for all three compounds [the dashed lines in Fig.~\ref{EFit}(a)]. After the interband contributions are removed, \Kratio\ saturates quickly with increasing $\omega_{c}$. Taking $\omega_{c}$ = 5000~\icm, we get \Kratio\ = 0.023 (\LNObi), 0.26 (\LNOtri) and 0.48 (\LNOinf).

Alternatively, \Kratio\ can also be determined from the Drude plasma frequency, as $K_{\text{exp}}/K_{\text{band}} = \omega_{p,exp}^{2}/\omega_{p,cal}^{2}$ with $\omega_{p,exp}$ and $\omega_{p,cal}$ corresponding to the experimental and calculated Drude plasma frequencies, respectively~\cite{SuppMat}. The Drude-Lorentz fit directly returns the values of $\omega_{p,exp}$ = 0.478, 1.85 and 2.66~eV for \LNObi, \LNOtri\ and \LNOinf, respectively. Note that $\omega_{p,exp}^{2} = \omega_{p,D1}^{2} + \omega_{p,D1}^{2}$, where $\omega_{p,D1}$ and $\omega_{p,D1}$ are the plasma frequencies of the two Drude components used in the Drude-Lorentz fit. The DFT calculations give $\omega_{p,cal}$ = 3.221, 3.53 and 3.7~eV for \LNObi, \LNOtri\ and \LNOinf, respectively. Consequently, we obtain \Kratio\ = 0.022 (\LNObi), 0.275 (\LNOtri) and 0.52 (\LNOinf), in good agreement with the values determined by integrating the experimental $\sigma_{1}(\omega)$.

\begin{figure}[tb]
\includegraphics[width=\columnwidth]{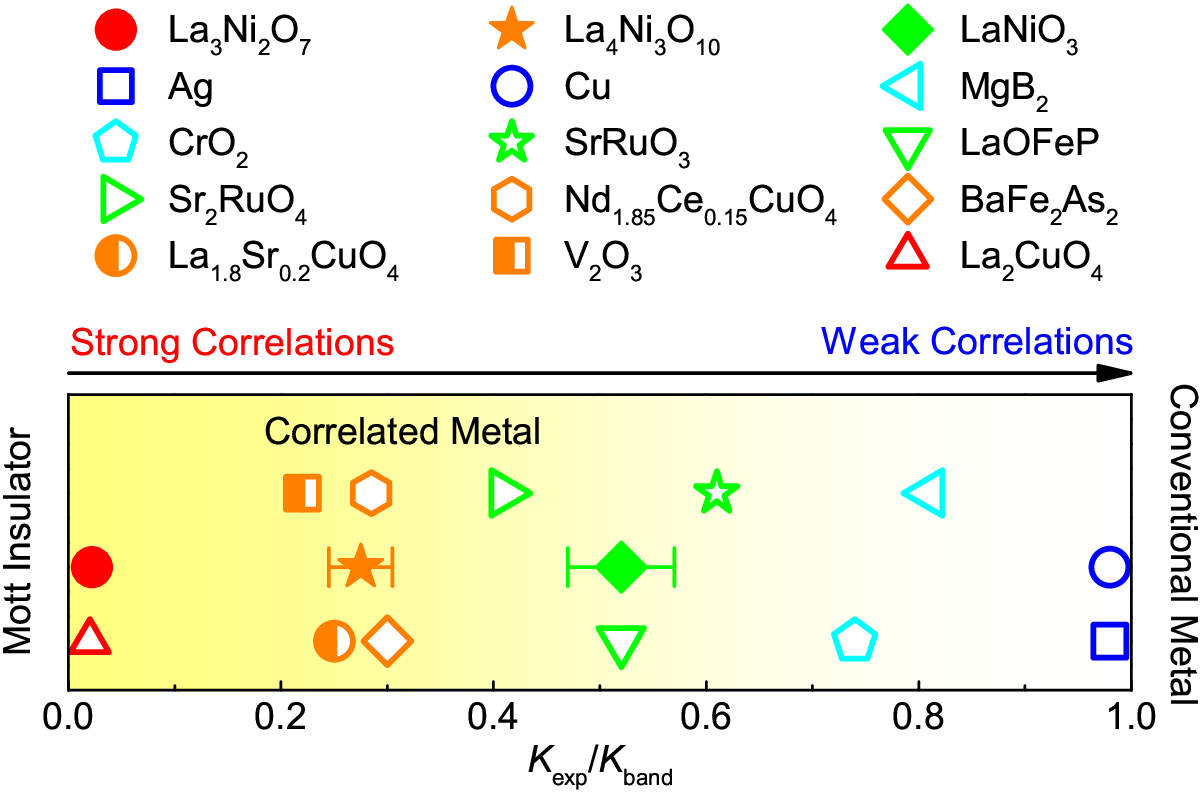}
\caption{(a) Kinetic energy ratio \Kratio\ for \LNObi\ (solid circle), \LNOtri\ (solid star), \LNOinf\ (solid diamond), and several other representative materials (open and half-filled symbols). The value of \Kratio\ for other materials are taken from Ref.~\cite{Qazilbash2009NP} and the references cited therein.}
\label{EEkRatio}
\end{figure}

Figure~\ref{EEkRatio} summarizes $K_{\text{exp}}/K_{\text{band}}$ for \LNObi\ (solid circle), \LNOtri\ (solid star), \LNOinf\ (solid diamond), and several other representative materials (open and half-filled symbols). In this diagram, conventional metals, for example Ag and Cu, are located in the regime of weak electronic correlations, where no substantial reduction of the electron's kinetic energy occurs, leading to a \Kratio\ close to unity. At the other end of the diagram lies the Mott insulators such as the parent compound of the high-$T_{c}$ cuprate superconductor La$_{2}$CuO$_{4}$ in which strong on-site Coulomb repulsion obstructs the motion of electrons, resulting in a significant reduction of $K_{\text{exp}}$ compared to $K_{\text{band}}$. Iron-based superconductors, e.g. LaOFeP and BaFe$_{2}$As$_{2}$, and doped cuprates such as La$_{1.8}$Sr$_{0.2}$CuO$_{4}$ and Nd$_{1.85}$Ce$_{0.15}$CuO$_{4}$ fall into the regime of correlated metal. These materials are characterized by \Kratio\ lying between those of conventional metals and Mott insulators, thus being categorized as moderately correlated materials. For the RP nickelates, while the value of $K_{\text{exp}}/K_{\text{band}}$ places the bilayer \LNObi\ on the verge of the Mott insulator phase, suggesting strong electronic correlations, the trilayer \LNOtri\ and infinite-layer \LNOinf\ are located in the correlated metal regime, resembling iron-based superconductors~\cite{Qazilbash2009NP} and doped high-$T_{c}$ cuprates~\cite{Millis2005PRB}. Hence the trilayer \LNOtri\ and infinite-layer \LNOinf\ belong to moderately correlated materials. Previous optical studies on \LNOinf\ thin films have obtained different \Kratio\ values ranging from 0.04 to 0.67 for different substrates~\cite{Ouellette2010PRB,Stewart2011PRB,Ardizzone2020PRB}, suggesting that the electronic correlations in the RP nickelates are very sensitive to strain which can effectively tune the lattice parameters.
%

\begin{figure}[tb]
\includegraphics[width=\columnwidth]{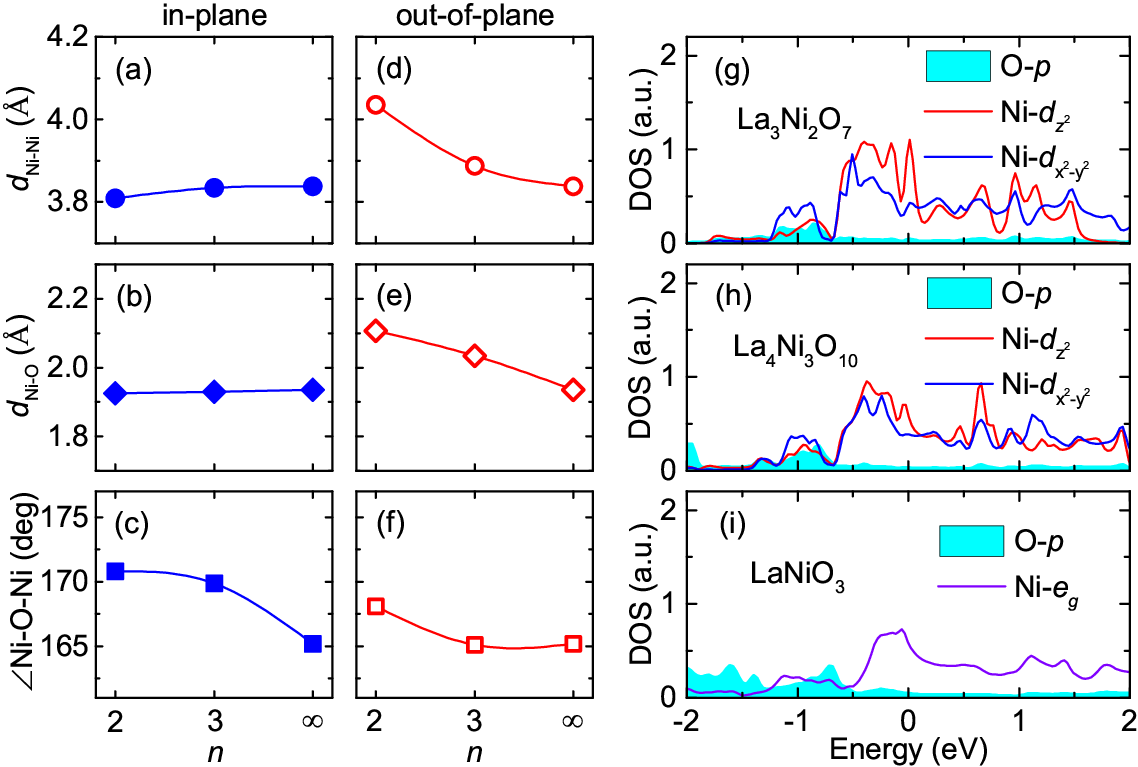}
\caption{(a)-(c) The evolution of the in-plane lattice parameters with $n$. (d)-(f) The evolution of the out-of-plane lattice parameters with $n$. (g)-(i) Calculated DOS for \LNObi, \LNOtri\ and \LNOinf.}
\label{EParvsn}
\end{figure}

%
Having established the evolution of electronic correlations in the RP nickelates, we next discuss its origin. The strength of electronic correlations in 3$d$ compounds depends sensitively on the lattice parameters which affect the electron hopping or hybridization between the $d$-$d$ or $p$-$d$ orbitals~\cite{Ouellette2010PRB,Stewart2011PRB,Ardizzone2020PRB,Zhu2010PRL,Yin2011NM,Nakajima2013PRB,Yi2015PRL,Zhou2023PRBCVNS}. For the RP nickelates, a smaller Ni-Ni distance ($d_{\text{Ni-Ni}}$) increases the hopping between the Ni-$3d$ orbitals; a shorter Ni-O bond length ($d_{\text{Ni-O}}$) and larger Ni-O-Ni angle ($\angle$Ni-O-Ni) enhance the hybridization between the Ni-$3d$ and O-$2p$ orbitals. All these effects reduce the electronic correlations. Figures~\ref{EParvsn}(a)-\ref{EParvsn}(c) trace out the in-plane $d_{\text{Ni-Ni}}^{\text{in}}$, $d_{\text{Ni-O}}^{\text{in}}$ and $\angle$Ni-O-Ni$^{\text{in}}$ for the three compounds~\cite{Chen2024JACS,Zhang2020PRM,Garcia-Munoz1992PRB}. As $n$ grows, while $d_{\text{Ni-Ni}}^{\text{in}}$ [Fig.~\ref{EParvsn}(a)] and $d_{\text{Ni-O}}^{\text{in}}$ [Fig.~\ref{EParvsn}(b)] increase slightly, $\angle$Ni-O-Ni$^{\text{in}}$ [Fig.~\ref{EParvsn}(c)] becomes smaller. These changes are clearly at odds with the reduction of electronic correlations, indicating that the in-plane lattice parameters are unlikely to play a decisive role in controlling the electronic correlations in the RP nickelates. The $n$ dependence of the out-of-plane $\angle$Ni-O-Ni$^{\text{out}}$ [Fig.~\ref{EParvsn}(f)] also conflicts with the evolution of electronic correlations, whereas the contraction of $d_{\text{Ni-Ni}}^{\text{out}}$ [Fig.~\ref{EParvsn}(d)] and $d_{\text{Ni-O}}^{\text{out}}$ [Fig.~\ref{EParvsn}(e)] are in accord with the electronic correlation reduction, suggesting that the out-of-plane Ni-Ni distance and Ni-O bond length act as one key factor to tune the electronic correlations in the RP nickelates. Since the in-plane and out-of-plane lattice parameters mainly affect the Ni-$d_{x^2-y^2}$ and Ni-$d_{z^2}$ orbitals, respectively, the evolution of the electronic correlations in the RP nickelates is most likely to be dominated by the Ni-$d_{z^2}$ orbital.

Another factor influencing the electronic correlations is the quantum confinement from the finite number of NiO$_{6}$ layers. In La$_{3}$Ni$_{2}$O$_{7}$, the spacing LaO layers block electron hoppings between NiO$_{6}$ bilayers. As a result, the two Ni-$d_{z^2}$ orbitals in the NiO$_{6}$ bilayer form inter-layer $\sigma$-bonding and antibonding bands that are essentially dispersionless along $c$ axis~\cite{Sun2023Nature,Yang2024NC,Liu2024NC}. The Ni-$d_{z^2}$ bonding band produces a van Hove singularity (vHS) sharply peaking near the Fermi level [Fig.~\ref{EParvsn}(g)], which is known to induce strong electronic correlations~\cite{Mravlje2011PRL,Li2010NP,Kugler2020PRL}. As the number of NiO$_{6}$ layers increases, the band dispersion along $c$ axis becomes pronounced, leading to a softening of the vHS in the Ni-$d_{z^2}$ partial DOS [Figs.~\ref{EParvsn}(h) and \ref{EParvsn}(i)]. This trend aligns with the observed reduction of electronic correlations with increasing $n$ and highlights the dominant role of the Ni-$d_{z^2}$ orbital in the evolution of electronic correlations in RP nickelates.

Finally, it is instructive to compare the evolution of electronic correlations and SC in the RP nickelates. Our optical studies have demonstrated that \LNObi\ features strong electronic correlations~\cite{Liu2024NC}, and the electronic correlations decrease with increasing $n$ in the RP nickelate family. Recent transport measurements under high pressure have found that while the maximum $T_{c}$ of \LNObi\ reaches 80~K~\cite{Sun2023Nature,Wang2024PRX,Zhang2024NP}, \LNOtri\ has a maximum $T_{c}$ of about 30~K~\cite{Zhu2024Nature,Li2024CPL}. Moreover, SC has not been reported in \LNOinf. These results seem to hint at a scaling relation between the electronic correlations and maximum $T_{c}$ in RP nickelates. Many theoretical models have emphasized the important role of strong correlations in driving SC in RP nickelates~\cite{Lechermann2023PRB,Liao2023PRB,Sakakibara2024PRL,Zhang2024PRL,Lu2024PRL,Yang2023PRB,Jiang2024CPL,Qu2024PRL,Oh2023PRB}, building on the localized nature of the Ni-$d_{z^2}$ orbitals. Our experimental findings of a correlation between reduced electronic correlations, especially in the Ni-$d_{z^2}$ orbitals, and the decrease in $T_{c}$ lend further credence to these studies.

%
%
To summarize, we determined the ratio of the kinetic energy from experimental $\sigma_{1}(\omega)$ and that from band theory $K_{\text{exp}}/K_{\text{band}}$ for the RP nicklates La$_{n+1}$Ni$_{n}$O$_{3n+1}$ with $n = 2$ (La$_{3}$Ni$_{2}$O$_{7}$), $n = 3$ (La$_{4}$Ni$_{3}$O$_{10}$) and $n = \infty$ (LaNiO$_{3}$). We found that $K_{\text{exp}}/K_{\text{band}}$ increases as $n$ grows, suggesting a reduction of electronic correlations with increasing NiO$_{6}$ layers. While the bilayer La$_{3}$Ni$_{2}$O$_{7}$ features strong electronic correlations with a vanishingly small $K_{\text{exp}}/K_{\text{band}}$ close to the Mott insulating phase, the trilayer La$_{4}$Ni$_{3}$O$_{10}$ and infinite-layer LaNiO$_{3}$ can be categorized as correlated metals with moderate electronic correlations. The change of electronic correlations in the RP nickelates is most likely to be dominated by the Ni-$d_{z^2}$ orbital.

\emph{Note added}--Recently, we became aware of an optical study~\cite{Xu2024arXiv} that found moderate electronic correlations in \LNOtri, consistent with our results.

%
%

\begin{acknowledgments}
We thank Ilya M. Eremin, Frank Lechermann, Qianghua Wang and Shunli Yu for helpful discussions. Work at NJU was supported by the National Key R\&D Program of China (Grants No. 2022YFA1403201 and 2022YFA1403000), the National Natural Science Foundation of China (Grants No. 12174180 and 12061131001). Work at SYSU was supported by the National Natural Science Foundation of China (Grants No. 12425404, 12474137 and 12174454), the National Key Research and Development Program of China (Grants No. 2023YFA1406000 and 2023YFA1406500), the Guangdong Basic and Applied Basic Research Funds (Grants No. 2024B1515020040 and 2024A1515030030), Shenzhen Science and Technology Program (Grant No. RCYX20231211090245050), and Guangdong Provincial Key Laboratory of Magnetoelectric Physics and Devices (Grant No. 2022B1212010008).
\end{acknowledgments}

%

\end{document}